# Large-area synthesis of continuous two-dimensional MoTe$_x$Se$_{2-x}$ alloy films by chemical vapor deposition

Shudong Zhao[1], Meilin Lu[1], Shasha Xue[1], Lei Tao[2], Yu Sui[1], Yang Wang[3]

[1] Department of Physics, Harbin Institute of Technology, Harbin 150001, People's Republic of China.

[2] Laboratory for Space Environment and Physical Sciences, Harbin Institute of Technology, Harbin 150001, People's Republic of China

[3] Academy of Fundamental and Interdisciplinary Sciences, Harbin Institute of Technology, Harbin 150001, People's Republic of China.

‡ Shudong Zhao and Meilin Lu contributed equally to this work

E-mail: taolei@hit.edu.cn (Lei Tao); suiyu@hit.edu.cn (Yu Sui); yangwang@hit.edu.cn (Yang Wang)

## Abstract

Great achievements have been made in alloying of two-dimensional (2D) semiconducting transition metal dichalcogenides (TMDs), which can allow tunable band gaps for practical applications in optoelectronic devices. However, telluride-based TMDs alloys were less studied due to the difficulties of sample synthesis. Here, in this work we report the large-area synthesis of 2D MoTe$_x$Se$_{2-x}$ alloy films with controllable Te composition by a modified alkali metal halides assisted chemical vapor deposition method. The as-prepared films have millimeter-scale transverse size. Raman spectra experiments combining calculated Raman spectra and vibrational images obtained by density functional theory (DFT) confirmed the 2H-phase of the MoTe$_x$Se$_{2-x}$ alloys. The A$_{1g}$ mode of MoSe$_2$ shows a significant downshift accompanied by asymmetric broadening to lower wavenumber with increasing value of $x$, while E$^1_{2g}$ mode seems unchanged, which were well explained by a phonon confinement model. Our work provides a simple method to synthesize large-scale 2H phase Te-based 2D TMDs alloys for their further applications.

# 1. Introduction

Two dimensional (2D) semiconducting transition metal dichalcogenides (TMDs) have shown great potential applications in the area of electronics[1], optoelectronics[2] and electrochemistry[3]. Alloying is a powerful method to tailor the properties of TMDs, which can tune their band gaps continuously between the two ends of one series as a function of chemical composition[4]. For instance, the direct-band gap of monolayer $MoS_xSe_{2-x}$ can be tuned between the values of single-layer $MoS_2$ (~1.9 eV) and $MoSe_2$ (~1.5 eV)[5-8]. In general there are two kinds of alloying in TMDs, one is at transition metal sites and another one is at chalcogenide site. Previously, several series of 2D semiconducting TMDs alloys have been obtained by exfoliation or chemical vapor deposition (CVD) methods, such as $Mo_{1-x}W_xS_2$[9, 10], $Mo_{1-x}W_xSe_2$[11], $WS_{2(1-x)}Se_{2x}$[12, 13]. These series of TMDs alloys usually have 2H structure, realizing the tunable direct-band gap from 1.5 eV to 2.1 eV.

It is difficult to prepared 2H-phase Te-based TMDs or their alloys, especially using CVD, because it always undergoes a phase transition from 2H to 1T' at cooling stage, thus resulting in the metallic or semi-metallic samples[14-16]. Amey Apte et al. reported a facile single-step CVD synthesis of 2H $MoTe_xSe_{2-x}$ monolayers[17]. However, the surface of as-grown nanosheets is not uniform and the size is small, limiting their applications in practice. Thus, a special method is pressingly needed to obtain 2D TMDs alloys at large scale.

Recently, molten-salt-assisted CVD has been successfully used to synthesize large-area and a wide variety of TMDs and their alloys[18, 19]. Here, we report a modified molten-salt-assisted CVD method to synthesize large area continuous 2D $MoTe_xSe_{2-x}$ films. Through tuning the heating temperature of Te powders, we prepared two kinds of $MoTe_xSe_{2-x}$ films with different Te alloying concentrations. Combining Raman scattering experiments and density functional theory calculation results we confirmed that the as-grown $MoTe_xSe_{2-x}$ films are 2H phase and the corresponding compositions of Te are about 0.11 and 0.25 respectively. The composition dependent behaviors of Raman peaks were well explained by a phonon confinement model. Our work provides a fundamental research for the applications of TMDs alloys.

# 2. Methods

## 2.1. Sample preparation

Large area continuous $MoTe_xSe_{2-x}$ films were synthesized in a home-built three-temperature-zone (a single-zone low-temperature furnace and a two-zone high temperature furnace) CVD system. The low-temperature furnace at upstream was used to heat Se powders (350 °C) and the high temperature furnace at downstream was used to heat Te powders (550 °C or 580 °C) and Mo precursors and substrates. The same weight (2 g) of Se and Te powders was used. 20 mg of $MoO_3$ powders and NaCl powders (2:1) were mixed by grinding together in a mortar and then were pressed into a disc at 10 MPa as the Mo precursors. The disc was placed on a quartz sheet and a piece of $SiO_2$/Si substrate was placed next to the disc at downstream. The pressure in the 2 inch furnace tube was firstly pumped down to 10 mTorr, then kept at 100 Torr with 55 sccm Ar and 3 sccm $H_2$ as the carrier gas. The growth temperature was 750 °C and time was 5 min. Then the furnace cooled down to room temperature gradually.

### 2.2. Raman characterization

Raman measurements were performed under atmospheric environment in a Renishaw inVia Raman microscope equipped with a 532 nm laser. The system was firstly calibrated with the strongest Raman peak of silicon at 520 $cm^{-1}$ before measurements.

### 2.3. Density functional theory calculation

Raman spectra and vibrational modes of monolayer $MoTe_xSe_{2-x}$ at two composition ($x = 0.11$ and $x = 0.25$) were calculated by density functional theory (DFT). The 3×3 supercell with one Te atom corresponds to $x = 0.11$, and the 2×2 supercell with one Te atom corresponds to $x = 0.25$. Vacuum spacing in the out-of-plane direction is set to be more than 18 Å to avoid spurious interaction. Structure and phonon properties are calculated by Vienna ab initio simulation package (VASP) using the projector-augmented wave (PAW) methods and Perdew-Burke-Ernzerhof (PBE) functional with D3 correction (DFT-D3). The energy cutoff was set as 400 eV and residual forces for ion iteration as $10^{-4}$ eV/Å. The Γ-centered Monkhorst-Pack (MP) k-points were set as 12×12×1 and 16×16×1 for $x = 0.11$ and $x = 0.25$ respectively. After the structural relaxations, the normal vibration modes are calculated within frozen phonon approximation. Giving these vibration mode, Raman intensities are obtained based on the calculation of the derivatives of the dielectric tensors with respect to phonon vibration using script "vasp_raman.py". This is a Raman off-resonant activity calculator using VASP as a back end. Lorentz broadening of the modes' intensities yielded the Raman spectrum.

## 3. Results and discussion

### 3.1. Synthesis of 2D $MoSe_{2(1-x)}Te_x$ film

Figure 1a shows the schematic diagram of our CVD process. We used the molten-salt-assisted CVD methods reported previously to synthesize 2D $MoSe_2$ nanosheets and $MoTe_xSe_{2-x}$ alloys[19]. At the pre-exploration stage of our experiments, the mixed $MoO_3$ and NaCl powders were used to grow $MoSe_2$ as previous works did (see Figure S1a). Contrary to our expectation, the size of as-prepared $MoSe_2$ turned out to be small (see Figure S1b). It seems to be ascribed to insufficient contact between $MoO_3$ and NaCl.

As we can see in Figure S1c, the mixed powders just partially reacted with Se through a short growth time (5 min). In order to make the $MoO_3$ and NaCl to contact sufficiently, we pressed the mixed powders to a disc as seen in Figure S1d. By this means, the full reaction of $MoO_3$, NaCl and Se was realized, resulting in large-area growth of $MoSe_2$ films (see Figure S1e and S1f).

Then this modified molten-salt-assisted method was used to synthesize 2D $MoSe_{2-x}Te_x$ alloy films. In a recent work, it has been demonstrated that large amount of Te alloying in $WSe_{2(1-x)}Te_{2x}$ (more than 0.5) leads to a phase transition from 2H to 1T'[20]. To avoid this phenomena, the heating temperature of Te powders was set at a lower point (550 °C). Figure 1(b) and 1(c) shows the optical images of as-prepared $MoSe_{2-x}Te_x$ continuous film at different magnifications, indicating the large area and its smooth surface. The scratch at the central area was made by sharp tweezers for the characterization of its thickness. As seen in figrue 1(d), the height image and profile of atomic force microscope (AFM) confirm that the $MoTe_xSe_{2-x}$ film is a bilayer and has a relatively smooth surface except a little amount of nanoparticles on it.

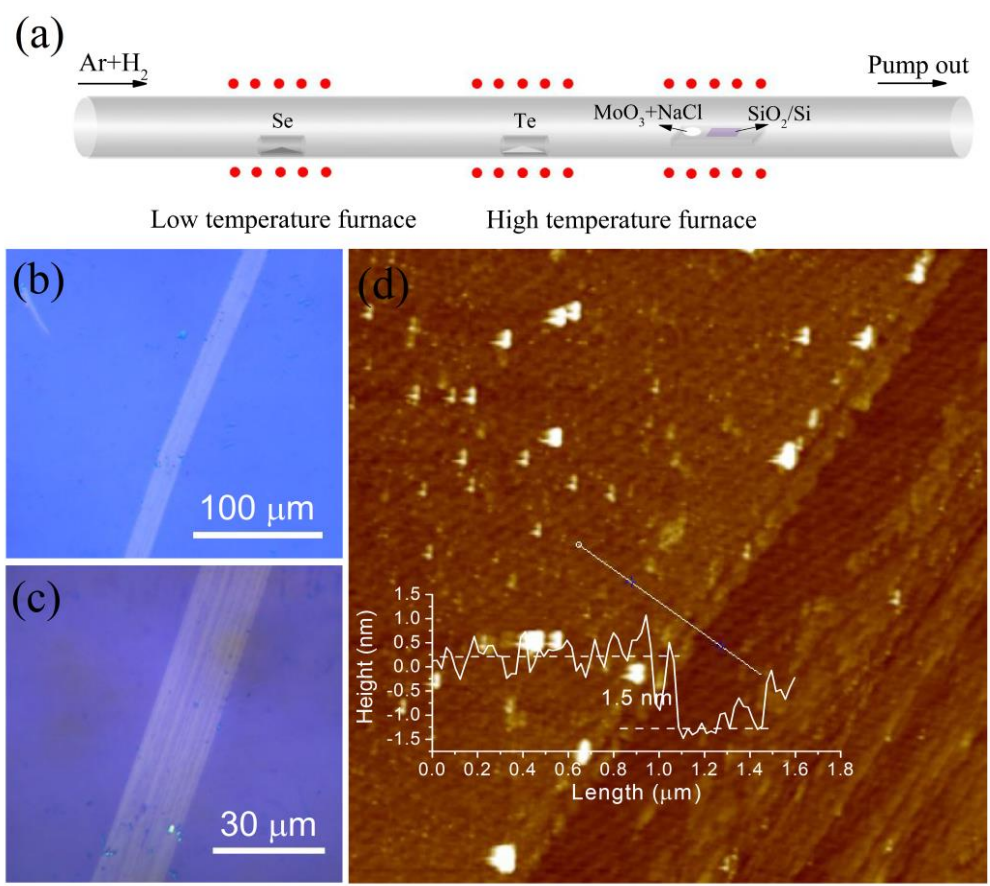

**Figure 1** (a) Schematic diagram of the CVD system. (b) Optical image of as-prepared $MoTe_xSe_{2-x}$ continuous film under 10x objective. (c) under 40x objective. (d) AFM height image and profile of the corresponding $MoTe_xSe_{2-x}$ film.

## 3.2. Raman characterization and the behaviors of Raman modes

The composition of Te in $MoTe_xSe_{2-x}$ alloys can be easily tailored by modifying the heating temperature of Te powders to 580 °C. Figure 2 gives the Raman spectra of the as-grown MoSe$_2$ and MoTe$_x$Se$_{2-x}$ alloys at two different Te heating temperatures. The Raman spectrum of pristine MoSe$_2$ shows one sharp peak at 240 cm$^{-1}$ corresponding to A$_{1g}$ mode, and a weak peak at 286 cm$^{-1}$ corresponding to E$^1_{2g}$ mode. With increasing composition $x$, the A$_{1g}$ peak shows a significant downshift (from 240 cm$^{-1}$ to 230 cm$^{-1}$) in MoTe$_x$Se$_{2-x}$ alloys, as well as the asymmetric broadening to lower frequency. However, the frequency of E$^1_{2g}$ mode seems unchanged and just shows a peak broadening.

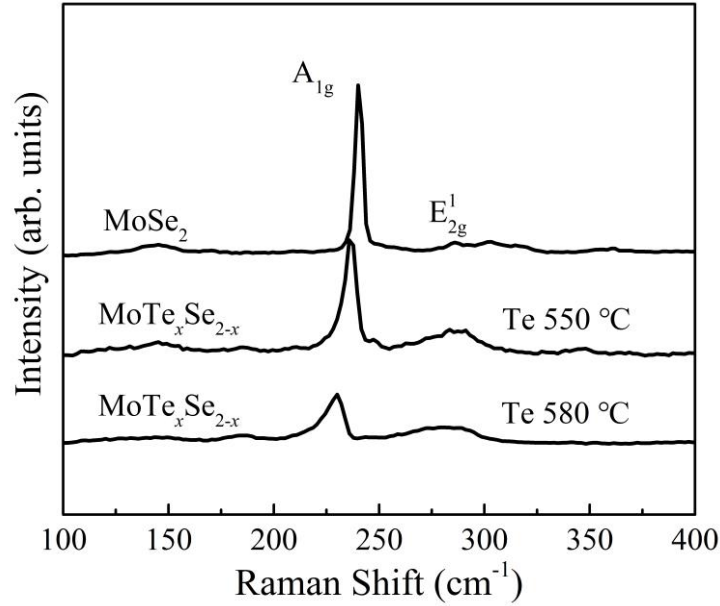

**Figure 2** Raman spectra of as-grown MoSe$_2$ and MoSe$_{2-x}$Te$_x$ alloys

Similar behaviors of A$_{1g}$ mode and E$^1_{2g}$ mode were also observed both in MoS$_x$Se$_{2-x}$ alloy series when $x \leqslant 1$[5] and in Se deficient MoSe$_{2-x}$ nanosheets[21]. The evolution of intrinsic Raman bands with increasing defect density can be explained by a phonon confinement model. After introducing defects in a crystalline material, the translational symmetry of the system is broken and the phonons are confined in a small region, resulting in relaxation of the fundamental Raman selection rule[22]. As to 2D system, the phonon confinement model can be described as[23]:

$$I(\omega) = \int \frac{\exp(-q^2 L_D/2\alpha)}{(\omega - \omega(q))^2 + (\Gamma_0/2)^2} 2\pi q \mathrm{d}q \qquad (1)$$

where $L_D$ is the average distance of defects or the domain size, corresponding to the density of defects and $\omega(q)$ is the phonon dispersion of 2D MoSe$_2$.

The phonon dispersion of monolayer MoSe$_2$ calculated by DFT is shown in figure 3(a). The dispersion curves of both two branches of E$^1_{2g}$ mode are relatively flat throughout the Brillouin zone. While, the frequency of A$_{1g}$ mode shows downshift along both Γ-K and Γ-M directions. The evolutions of positions of E$^1_{2g}$ and A$_{1g}$ peaks in Raman spectra agree very well with their phonon dispersion. Because the E$^1_{2g}$ peak is weak and its peak position changes little, we just simulated the Raman spectra of $A_{1g}$ mode using equation (1). The simulation results are illustrated in figure 3(b) and 3(c), the value of $\alpha$ we used is 0.001. As can be seen, the simulated curves fit well with the experimental data.

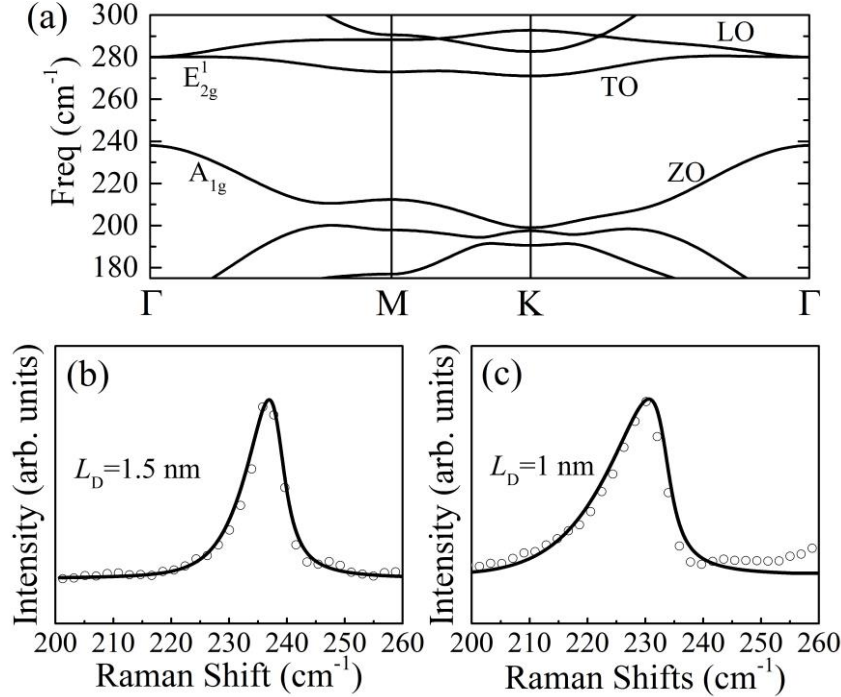

**Figure 3** (a) Phonon dispersion of monolayer $MoSe_2$. (b) and (c) Simulated Raman spectra of 2D $MoTe_xSe_{2-x}$ films, black line is the simulation result and black circle is the experimental data.

In addition, we also used DFT method calculated the vibrational modes and Raman spectra of 2H phase $MoTe_xSe_{2-x}$ at two Te alloying concentrations, 0, 0.11 and 0.25 respectively (see details in Methods and figure S2). As illustrated in figure 4, when $x = 0$, it is $MoSe_2$ and the modes at 238 $cm^{-1}$ and 280 $cm^{-1}$ represent $A_{1g}$ and $E^1_{2g}$ modes respectively. Se atoms at top and bottom layers vibrate out phase along vertical direction while Mo atoms are static in $A_{1g}$ mode. As in $E^1_{2g}$ mode, all Se atoms vibrate in phase as well as Mo atoms along armchair or zigzag directions (LO or TO), but Se and Mo atoms vibrate out phase. These are typical vibrational images of optical phonons of monolayer $MoSe_2$. When $x = 0.11$, there are also two mode at 235 $cm^{-1}$ and at 280 $cm^{-1}$. In general, the vibrational image of mode at 235 $cm^{-1}$ is quite similar with $A_{1g}$ mode in $MoSe_2$. That is atoms in top chalcogenide layer still vibrate in phase along vertical direction and out phase with the atoms in bottom layer, so it can be assigned to $A_{1g}$ mode. As to the mode at 280 $cm^{-1}$, all chalcogenide atoms vibrate in phase and out phase with Mo atoms in $xy$ plane, so it can be assigned to $E^1_{2g}$ mode. The same phenomena can be observed when $x = 0.25$, but just the $A_{1g}$ mode moves to 230 $cm^{-1}$ and the frequency of $E^1_{2g}$ mode is unchanged. The evolution of peak frequency shifts of the first-order Raman modes obtained by DFT calculation shows high agreement with the experimental results. And surprisingly, the calculated peak positions of MoTe$_x$Se$_{2-x}$ are just corresponding to the positions in the experimental Raman spectra, which can be an indirect characterization of the Te alloying concentrations in MoTe$_x$Se$_{2-x}$ films.

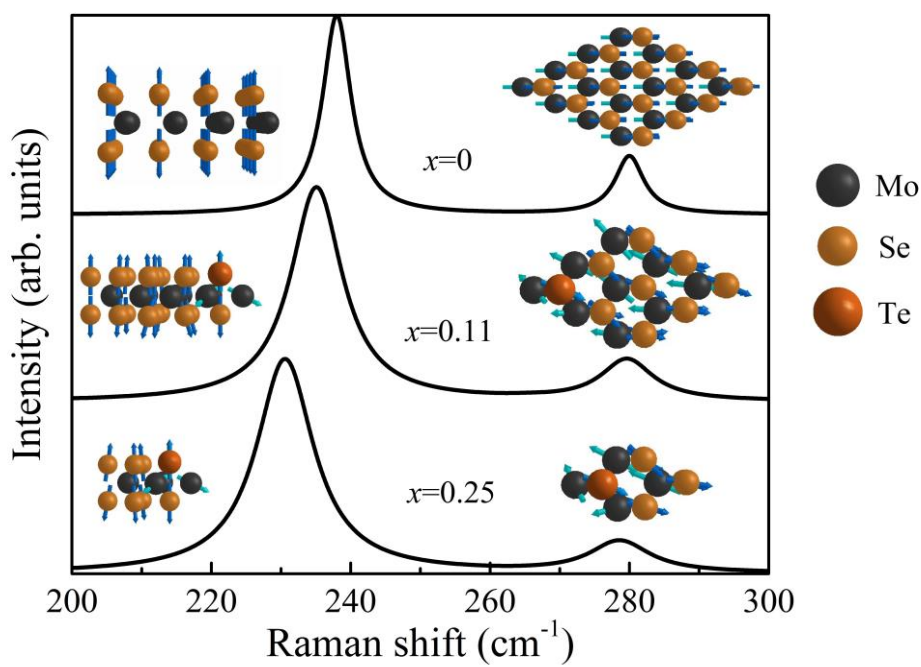

**Figure 4** Calculated Raman spectra and atomic vibrational images of MoTe$_x$Se$_{2-x}$ films at three different compositions.

## 4. Conclusions

In summary, we have synthesized large-area continuous 2H phase MoSe$_{2-x}$Te$_x$ films, and tailored the Te composition through modifying the heating temperature of Te powders independently in a three-temperature-zone CVD process. The evolution of peak position shifts and linewidths of the A$_{1g}$ mode and E$^1_{2g}$ mode is well explained with the phonon confinement model using the phonon dispersion of MoSe$_2$ calculated by DFT. On the other hand, the Raman spectra and atomic vibrational images of 2D MoSe$_{2-x}$Te$_x$ obtained by DFT calculation give an additional view of the behaviors of the first-order optical phonons with increasing Te compositions, which also agree very well with the experimental data. This work provides a simple method to synthesize large-scale 2D MoTe$_x$Se$_{2-x}$ alloy series and gives a fundamental study of their Raman scattering properties.

## Supporting information

Supporting information is available on line or from author.

## Acknowledgements

This study was financially supported by the National Natural Science Foundation of China (No. 11304060) and the Foundation of Harbin Institute of Technology for the

# Supporting Information

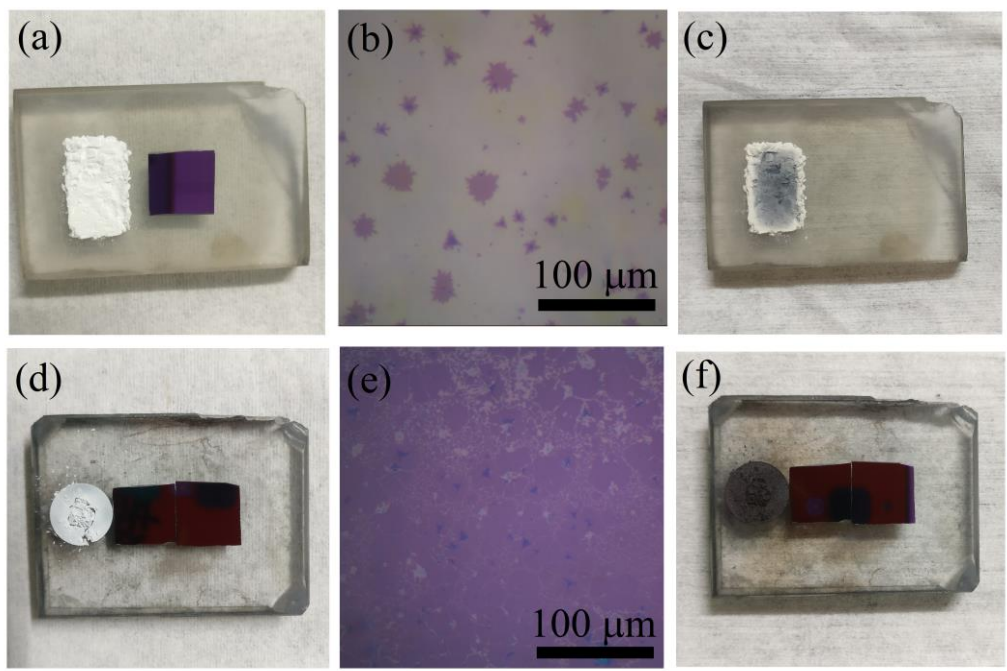

**Figure S1** Comparations of the traditional NaCl assisted CVD method and our modified NaCl assisted CVD method

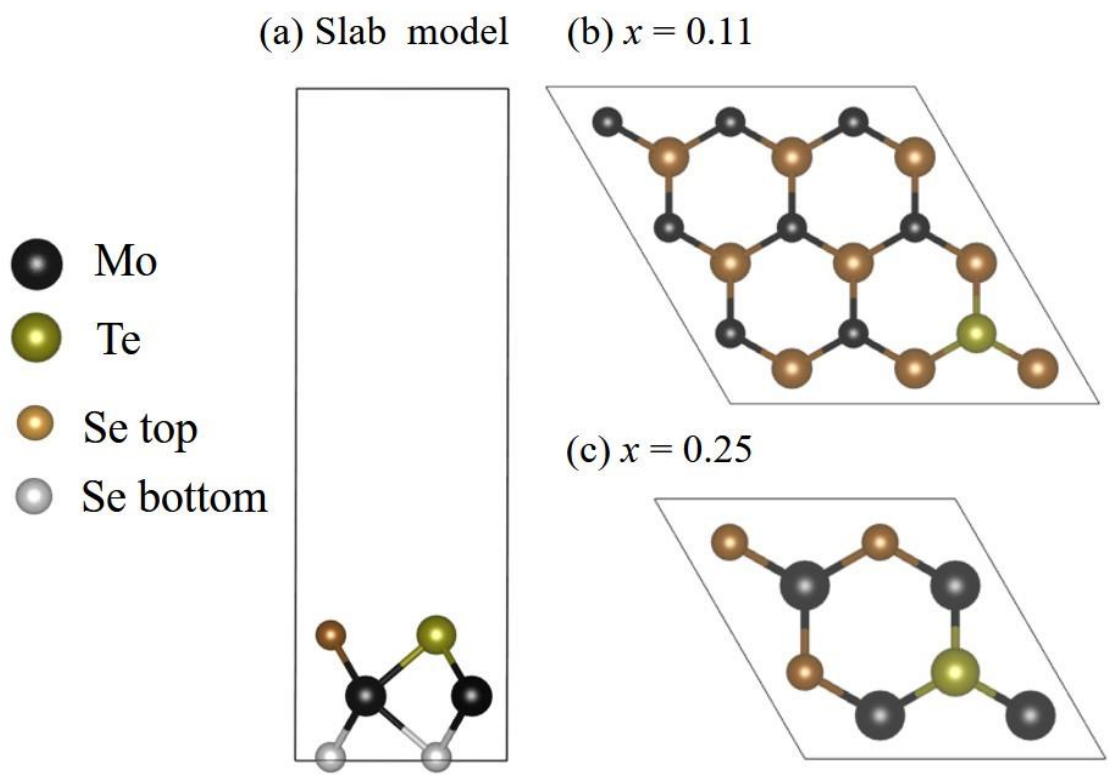

**Figure S2** Slab models of 2D MoTe$_x$Se$_{2-x}$ used in our DFT calculation